\documentclass[journal]{IEEEtran} 

\usepackage{bm}
\usepackage{amsmath}
\usepackage{epsf}
\usepackage{graphics}
\usepackage{ amssymb }
\usepackage[dvips]{graphicx}
\usepackage{mathtools}
\usepackage{epsfig}
\usepackage{cite}
\usepackage[linesnumbered,ruled,vlined]{algorithm2e}
\usepackage{colortbl}
\usepackage{color}
\usepackage{hyperref} 
\usepackage{enumitem}
\usepackage{soul,xcolor}

\usepackage{bm}
\usepackage{amsmath}
\usepackage{epsf}
\usepackage{graphics}
\usepackage{ amssymb }
\usepackage[dvips]{graphicx}
\usepackage{epsfig}
\usepackage{cite}
\usepackage[linesnumbered,ruled,vlined]{algorithm2e}
\usepackage{graphicx}
\usepackage{epsfig}
\usepackage{latexsym}
\usepackage{amsfonts}
\usepackage{here}
\usepackage{rawfonts}
\usepackage[utf8]{inputenc}
\usepackage[english]{babel}
\usepackage{amsmath}
\usepackage{amsfonts}
\usepackage{amssymb}
\usepackage{color}
\usepackage{bm}
\usepackage{listings}
\usepackage{caption}
\usepackage{multicol}
\usepackage{tabularx,booktabs}
\newcolumntype{Y}{>{\centering\arraybackslash}X}
\makeatletter
\newcommand\notsotiny{\@setfontsize\notsotiny{6.31415}{7.1828}}
\makeatother
\usepackage{amssymb}
\usepackage{amsthm}
\usepackage{graphicx}
\usepackage{epstopdf}
\usepackage{listings}
\usepackage{float}
\usepackage{amsmath}
\usepackage{amssymb}
\usepackage{amsfonts}
\usepackage{epstopdf}

\usepackage{multirow}
\usepackage{amscd}
\usepackage{mathrsfs}
\usepackage{graphicx}
\usepackage{color}
\usepackage{url}
\usepackage{bm}
\usepackage{setspace}
\usepackage{footnote}
\usepackage{xcolor}
\definecolor{lightblue}{RGB}{173, 216, 230}  
\newtheorem{remark}{Remark}

\usepackage[]{algorithm2e}

\addto\captionsenglish{}
\usepackage{bbm}

\usepackage{multicol}
\usepackage{mathtools}

\usepackage{lipsum,graphicx,subcaption}

\usepackage{diagbox}
\usepackage{hyperref}
\usepackage[protrusion=true,expansion=true]{microtype}
\pdfoutput=1
\usepackage[font=footnotesize]{caption}
\allowdisplaybreaks[4]

\let\emptyset\varnothing

\usepackage{graphicx}
\usepackage{grffile}
\usepackage{tabularx}
\newcounter{term}[section]

\renewcommand\theterm{\alph{term}}
\makeatletter
\newcommand{\vast}{\bBigg@{4}}

\newcommand{\Vast}{\bBigg@{5}}
\makeatother
\newcommand\semiHuge{\fontsize{22.7}{31.38}\selectfont}
\usepackage{soul,xcolor}

\usepackage[most]{tcolorbox}
\tcbset{colback=yellow!10!white, colframe=black!50!black, 
        highlight math style= {enhanced, 
            colframe=black,colback=red!10!white,boxsep=0pt}
        }
        \allowdisplaybreaks

\begin{document} 
\title{{\semiHuge Cross-Layer Traffic Allocation and Contention
Window Optimization for Wi-Fi 7 MLO: When DRL Meets LSTM}}
\author{Zhang Liu, Xianbin Wang,~\IEEEmembership{Fellow,~IEEE}, Shumin Lian, Lianfen Huang,\\ Liqun Fu,~\IEEEmembership{Senior Member,~IEEE}, and Ying-Jun Angela Zhang,~\IEEEmembership{Fellow,~IEEE}

\thanks{\emph{Z. Liu (zhangliu@xmu.edu.cn, zhangliu@cuhk.edu.hk) is with the Department of Computer Science and Technology, Xiamen University, China, and also with the Department of Information Engineering, The Chinese University of Hong Kong, Hong Kong. X. Wang (xianbin.wang@uwo.ca) is with the Department of Electrical and Computer Engineering, Western University, Canada. L. Huang (lfhuang@xmu.edu.cn) is with the Key Laboratory of Intelligent Manufacturing Equipment and Industrial Internet Technology, School of Information Science and Technology, Xiamen University Tan Kah Kee College, China, and also with the Department of Informatics and Communication Engineering, Xiamen University, China. S. Lian (smlian@stu.xmu.edu.cn) and L. Fu (liqun@xmu.edu.cn) are with the Department of Informatics and Communication Engineering, Xiamen University, China. Y. Zhang (yjzhang@ie.cuhk.edu.hk) is with the Department of Information Engineering, The Chinese University of Hong Kong, Hong Kong. (Corresponding author: Xianbin Wang.)} }
} 
\maketitle
\vspace{-9mm}
\setulcolor{red}
\setul{red}{2pt}
\setstcolor{red}   

\begin{abstract}
To support future diverse applications, multi-link operation (MLO) has been introduced in the Wi-Fi 7 standard (IEEE 802.11be) to enable concurrent communication over multiple frequency bands. This new capability relies on a two-tier medium access control (MAC) architecture, where the upper MAC (U-MAC) allocates traffic across links and the lower MAC (L-MAC) performs independent channel access. {However, MLO optimization is challenging due to the inherent coupling between the U-MAC and L-MAC, as well as the dynamic and complex nature of wireless networks. To address these challenges, we propose a cross-layer framework that jointly optimizes traffic allocation at the U-MAC layer and initial contention window (ICW) sizes at the L-MAC layer to maximize network throughput.} Specifically, we extend the single-link Bianchi Markov model to develop an analytical framework that captures the relationship among network throughput, traffic allocation, and ICW sizes. Based on this framework, we formulate a nonconvex, nonlinear cross-layer optimization problem. To solve it efficiently, we design a long short-term memory–based soft actor–critic (LSTM-SAC) algorithm that leverages LSTM to handle the partial observability and non-Markovian dynamics inherent in Wi-Fi networks. Finally, using a well-developed event-based Wi-Fi simulator, we demonstrate that the proposed LSTM-SAC substantially outperforms existing benchmark solutions across a wide range of network settings.


\end{abstract}
\vspace{-3mm}
\section{Introduction} \label{sec:intro}
\subsection{Background and Overview} \label{subsec:background}
\vspace{-.15mm}
Over more than three decades of rapid evolution, Wi-Fi has become the world’s most widely used indoor wireless networking technology. Owing to its ease of deployment, flexibility, and cost-effectiveness, Wi-Fi networks now carry approximately 70\% of global Internet traffic~\cite{lian2025intelligent}. To support increasingly diverse future services and applications, several new features have been introduced at both the physical and medium access control (MAC) layers in the IEEE 802.11be Extremely High Throughput (EHT) amendment~\cite{IEEE80211beD7}, also known as Wi-Fi 7. At the physical layer, Wi-Fi 7 supports increased channel bandwidths of up to 320 MHz, enabling extremely high data rates and low-latency communication. At the MAC layer, it introduces multiple resource unit capabilities and multi-link operation (MLO)\cite{iturria2023rl}. Together, these features provide substantial throughput gains and support emerging applications such as virtual and augmented reality, remote collaboration and teleoperation, and cloud gaming\cite{deng2020ieee}.

\begin{figure}[t!]
\includegraphics[width=.48\textwidth]{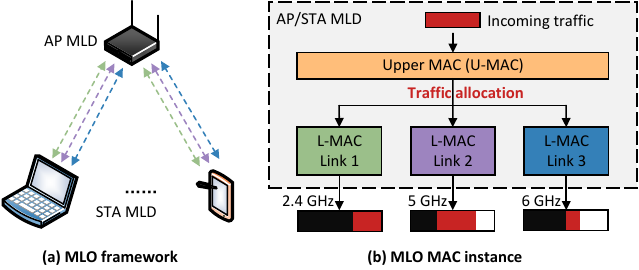}
\centering
\vspace{-1.5mm}
\caption{A schematic illustration of the MAC instance in MLO, where each L-MAC in a different color represents a distinct band corresponding to one of the multiple links. In addition, the traffic allocated to each link is highlighted in red, while the channel occupancy is shown in black.}
\label{fig:MLO MAC}
\end{figure}

Among these new features, MLO -- the focus of this work -- is a new foundational MAC layer technique that enables access points (APs) and stations (STAs) to perform concurrent transmissions over multiple links operating across different frequency bands, including 2.4/5/6 GHz. To this end, multi-link devices (MLDs), including AP MLDs and STA MLDs,\footnote{The APs and STAs with MLO capability are defined as MLDs, referred to as AP MLDs and STA MLDs, respectively. In this paper, we assume that all APs and STAs are capable of MLO, without considering coexistence scenarios with legacy Wi-Fi devices. For simplicity, we refer to AP MLDs and STA MLDs as APs and STAs throughout the remainder of the paper.} provide a unique MAC instance bifurcated into two MAC sublayers~\cite{lopez2022multi}, each with distinct functionalities. As shown in Fig.~\ref{fig:MLO MAC}(b), the upper MAC (U-MAC) serves as a common sublayer for all links and is responsible for the aggregation and de-aggregation of MAC service data units. In contrast, the lower MAC (L-MAC) handles link-specific functions, operating independent channel access for each link. {Although the two-layer MAC architecture enables flexible multi-link transmission, it also poses significant optimization challenges. In particular, the overall performance of MLO is jointly determined by traffic allocation at the U-MAC and link-specific channel access at the L-MAC, which are inherently interdependent. This cross-layer coupling, further exacerbated by the dynamic nature of wireless channel and data traffic, makes efficient MLO optimization a highly nontrivial task.}

\subsection{Motivation and Main Challenges} \label{subsec:challenges}
\vspace{-.15mm}

To improve MLO performance, several traffic allocation policies at the U-MAC layer have been proposed~\cite{alsakati2023performance, lopez2021ieee, gao2025latency, lopez2022dynamic} (as detailed in Sec.~\ref{subsec:traffictolink}). {\emph{\textbf{However, these studies focus solely on traffic allocation at the U-MAC layer and do not account for channel contention at the L-MAC layer.}} In other words, they implicitly assume that all links can access the channel within a reasonable time and successfully transmit their allocated traffic.} In practice, Wi-Fi adopts carrier-sense multiple access with collision avoidance (CSMA/CA), where each AP/STA performs a random backoff before attempting channel access, thereby introducing significant uncertainty (as detailed in Sec.~\ref{subsec:channelaccess}). {To address this issue, we explicitly incorporate the impact of channel access parameters -- namely, the \emph{{initial contention window (ICW)}} sizes at the L-MAC layer -- and propose a cross-layer framework that jointly optimizes traffic allocation and ICW sizes to maximize network throughput.} To this end, we extend the widely adopted Bianchi Markov model for single-link Wi-Fi~\cite{bianchi2002performance} to derive explicit expressions that relate network throughput to both the traffic allocation policy and the ICW size in Wi-Fi 7 MLO (as detailed in Sec.~\ref{subsec:analyticalframework}). This analysis reveals that network throughput is jointly determined by two key factors: (i) the traffic allocation policy at the U-MAC layer and (ii) the ICW size at the L-MAC layer, thereby providing a solid theoretical foundation for the subsequent optimization.

Most existing studies employ analytical approaches~\cite{gao2025latency, dai2012unified, park2024adaptive, 10845889} to derive closed-form traffic allocation policies or ICW sizes. {\emph{\textbf{However, these analytical methods typically rely on quasi-static assumptions, which limit their applicability in realistic Wi-Fi 7 MLO scenarios.}} This limitation stems from the fact that closed-form solutions often scale poorly with network size and lack adaptability to time-varying traffic patterns and dynamic contention conditions.} Recently, the research community has increasingly turned to artificial intelligence (AI) to optimize Wi-Fi performance~\cite{wilhelmi2024machine}, with deep reinforcement learning (DRL) emerging as a particularly promising approach. In DRL, an intelligent agent (e.g., the AP) observes the current environment state (e.g., channel occupancy), selects an action (e.g., a traffic allocation policy), and receives a reward signal (e.g., network throughput). Through this trial-and-error interaction, the agent learns adaptive strategies suited to dynamic environments. {Motivated by these advantages, we adopt the soft actor–critic (SAC) algorithm~\cite{haarnoja2018soft} to jointly optimize traffic allocation and ICW sizes.} SAC is based on the maximum-entropy framework, which aims to maximize not only the expected cumulative reward but also the policy entropy. This entropy-regularized objective encourages persistent exploration and mitigates premature convergence to suboptimal solutions.

\begin{figure}[t!]
\includegraphics[width=.4\textwidth]{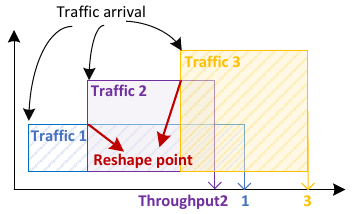}
\centering
\vspace{-1.5mm}
\caption{Non-Markovian nature of Wi-Fi networks: The throughput of the current traffic flow depends not only on the present state but is also influenced by subsequent flow arrivals.}
\label{fig:nonMarkovian}
\end{figure}

\emph{\textbf{Despite its promising potential, direct application of DRL to Wi-Fi 7 MLO networks poses substantial challenges.}} First, decision-making in Wi-Fi networks is inherently partially observable. A DRL agent (e.g., the AP) cannot directly access fine-grained MAC-layer states, such as the exact backoff counters of competing STAs or instantaneous collision events on each link. Consequently, the underlying optimization problem is more accurately modeled as a partially observable Markov decision process (POMDP)~\cite{wydmanski2021contention}.\footnote{Unlike standard MDPs, the true system state in a POMDP is not directly accessible. The agent must therefore approximate hidden states, making policy learning more difficult and less stable.} Second, the system exhibits non-Markovian behavior.\footnote{The Markov property assumes that the future state of an environment depends only on its current state and the action taken.} As illustrated in Fig.~\ref{fig:nonMarkovian}, the throughput resulting from a given traffic allocation policy and ICW size for Traffic~1 is strongly affected by subsequent actions, since later traffic arrivals dynamically reshape channel congestion and contention levels~\cite{iturria2023rl}. Together, these two factors significantly degrade the effectiveness and training stability of conventional DRL methods. {To address these challenges, we incorporate long short-term memory (LSTM) into SAC and propose an LSTM-SAC algorithm. The LSTM network encodes a history of past observations into a compact latent representation that approximates the underlying hidden system state (as detailed in Sec.~\ref{subsec:LSTM}). This capability allows the DRL agent to capture temporal correlations induced by non-Markovian dynamics and to infer unobservable network conditions under partial observability, thereby enabling more stable training.}

\vspace{-3mm}
\subsection{Summary of Contributions} \label{subsec:contributions}
\vspace{-.15mm}

\emph{\textbf{To the best of our knowledge, this is the first work on Wi-Fi 7 MLO that jointly incorporates cross-layer optimization by addressing traffic allocation at the U-MAC layer and ICW optimization at the L-MAC layer.}} Our main contributions are as follows:

\begin{enumerate}[leftmargin=4mm]
\item \textbf{\textit{Modeling:}} We extend the Bianchi Markov model from the single-link case to develop a novel analytical framework for characterizing the network throughput of Wi-Fi 7 MLO. Within this framework, we derive explicit expressions for network throughput as functions of the traffic allocation policy at the U-MAC layer and the ICW size at the L-MAC layer. Building on this analysis, we formulate a cross-layer optimization problem, which is shown to be non-convex and nonlinear through rigorous theoretical examination.

\item \textbf{\textit{Solution:}} To address this problem effectively, we propose an LSTM-SAC algorithm to jointly optimize the traffic allocation policy and the ICW size. Specifically, built upon the maximum-entropy DRL framework, SAC encourages persistent exploration and mitigates premature convergence to suboptimal solutions. LSTM networks are then incorporated to cope with the partial observability and non-Markovian behavior inherent in Wi-Fi networks, where a history of past observations is fed into the LSTM to produce a compact latent representation, which is subsequently used as the system state for training the DRL agent.

\item \textbf{\textit{Validation:}} We implemented the LSTM-SAC algorithm in \emph{Python} using OpenAI Gym for DRL-based learning, and integrated it with a well-established event-based Wi-Fi simulator~\cite{lian2025intelligent,tong2021throughput} implemented in \emph{Matlab}. The two platforms interact iteratively during training and evaluation. Simulation results demonstrate that the traffic allocation policies and ICW sizes obtained by the proposed LSTM-SAC algorithm consistently outperform existing benchmark solutions across a wide range of network settings.
\end{enumerate}

\vspace{-3mm}
\subsection{Paper Organization} \label{subsec:organization}
\vspace{-.15mm}
The rest of this paper is organized as follows. Sec.~\ref{sec:relatedwork} reviews related work. Sec.~\ref{sec:systemmodel} describes the Wi-Fi~7 MLO system model and formulates the corresponding cross-layer optimization problem. Sec.~\ref{sec:LSTM} introduces the preliminaries of LSTM networks. Sec.~\ref{sec:LSTMSAC} presents the proposed LSTM-SAC algorithm. Sec.~\ref{sec:simulation} reports and discusses the simulation results, followed by conclusions and future research directions in Sec.~\ref{sec:conclusion}.

\vspace{-3mm}
\section{Related Work} \label{sec:relatedwork}
\vspace{-.15mm}
Henceforth, we summarize the contributions of related works and highlight the aspects they have not addressed, which serve as the primary motivations for this work.

\vspace{-3mm}
\subsection{Traffic Allocation in Wi-Fi 7 MLO}\label{subsec:traffictolink}
\vspace{-.15mm}

When using MLO, traffic flows at the transmitter are distributed across different links according to an implementation-specific traffic allocation policy. The authors in~\cite{alsakati2023performance} proposed a condition-aware policy that considers the transmission data rate of each link, thereby maximizing the number of STAs that Wi-Fi can support. The authors in~\cite{lopez2021ieee} studied a congestion-aware policy which distributes incoming traffic according to the channel occupancy, with the goal of maximizing spectrum utilization. The authors in~\cite{gao2025latency} developed an analytical model for the latency performance of coexisting Wi-Fi 7 networks and numerically derived optimal traffic allocation strategies for minimizing mean end-to-end delay and delay jitter. The authors in~\cite{lopez2022dynamic} introduced a dynamic congestion-aware policy which periodically adjusts traffic allocation in response to spectrum occupancy changes, aiming to minimize the negative impact from neighboring APs. 

However, the above studies focus solely on the traffic allocation policy at the U-MAC layer, overlooking cross-layer interdependencies (e.g., the influence of the ICW size at the L-MAC layer). Intuitively, ICW optimization directly determines each link’s channel access effectiveness, which fundamentally shapes the actual gains achievable through traffic allocation. 

\vspace{-3mm}
\subsection{Contention Window Optimization in Wi-Fi 7 MLO}\label{subsec:CWoptimization}
\vspace{-.15mm}
ICW optimization plays a crucial role in Wi-Fi networks, as each node (including APs and STAs) performs a random backoff before attempting to access the channel. The authors in~\cite{park2024adaptive} obtained appropriate ICW sizes for STAs in coexisting Wi-Fi networks, enabling each STA to access the channel fairly alongside legacy STAs (i.e., STAs without MLO capabilities). The authors in~\cite{zhang2022synchronous} derived explicit expressions for the maximum network sum rate and the corresponding optimal ICW sizes, aiming to maximize the network throughput. The authors in~\cite{lv2025multi} proposed exponentially increasing and decreasing ICW schemes for STAs by allowing them to alternate between different ICW update rules, thereby enabling low-latency wireless local area networks. The authors in~\cite{korolev2022study} evaluated various channel access schemes under non-simultaneous transmit and receive operation, with the goal of achieving efficient and fair channel access in congested Wi-Fi~7 MLO networks.

However, the above studies are based on analytical models with restrictive assumptions. They typically presume quasi-static network conditions and simplified contention behaviors, and thus fail to capture dynamic traffic variations, cross-link interactions, and the strong coupling between traffic allocation and channel access inherent in Wi-Fi~7 MLO.

\vspace{-3mm}
\subsection{Usage of Deep Reinforcement Learning in Wi-Fi}\label{subsec:DRLWiFi}
\vspace{-.15mm}
The IEEE 802.11 working group is exploring the use of DRL to develop intelligent strategies for improving the performance of Wi-Fi networks. The authors in~\cite{lian2025intelligent} modeled the channel allocation problem within a multi-armed bandit framework to enable online learning for maximizing network throughput. The authors in~\cite{wydmanski2021contention} employed two DRL algorithms -- deep Q-network and deep deterministic policy gradient -- to learn optimal ICW sizes under varying network conditions, with the goal of maximizing network throughput. The authors in~\cite{tai2024model} developed a DRL approach based on SAC to dynamically determine flexible traffic allocation policies, achieving improved average network throughput. The authors in~\cite{ali2018deep} proposed an intelligent Q-learning–based resource allocation mechanism for MAC-layer channel access, aiming to maximize network throughput while minimizing channel access delay.

However, the above studies mostly rely on conventional DRL algorithms, which fail to address the partial observability and non-Markovian behavior inherent in Wi-Fi networks. These two factors significantly degrade the effectiveness and training stability of conventional DRL methods, highlighting the need for further algorithmic development.

\vspace{-3mm}
\section{System Model and Problem Formulation} \label{sec:systemmodel}
\vspace{-.15mm}
In this section, we first revisit the IEEE 802.11 channel access rules and provide an overview of Wi-Fi 7 MLO. We then derive network throughput as an explicit function of the traffic allocation policy and ICW sizes. Finally, we formulate the corresponding cross-layer optimization problem.

\vspace{-3mm}
\subsection{Carrier-Sense Multiple Access with Collision Avoidance}\label{subsec:channelaccess}
\vspace{-.15mm}

The enhanced distributed channel access (EDCA) mechanism generalizes the basic distributed coordination function (DCF) by allowing each STA to use its own medium-access parameters. Unlike DCF -- where all devices share identical ICW sizes -- EDCA enables each STA to operate with distinct ICW, thereby providing differentiated channel-access behavior. Specifically, under EDCA, once an STA senses that the channel is idle for an arbitration inter-frame space (AIFS), it initializes its backoff counter with a random value uniformly drawn from $[0, w-1]$, where $w$ is the CW size determined by its previous transmission failures. At the first transmission attempt, $w$ is set to the STA’s ICW $w_0$. After each unsuccessful attempt, $w$ doubles until reaching $2^M w_{0}$, where $M$ is the maximum backoff stage.\footnote{The contention window (CW) evolves dynamically according to the exponential backoff rule after each failed transmission, while the initial contention window (ICW) specifies the starting CW value before any retransmissions. In this paper, we focus on optimizing the ICW, as it directly determines the baseline channel access aggressiveness.} The backoff counter decreases while the channel is idle, freezes when the medium becomes busy, and resumes after the channel is sensed idle for at least an AIFS. The STA transmits when the backoff counter reaches zero.\footnote{We briefly summarize the CSMA/CA mechanism defined in the 802.11 protocol, a complete description can be found in the standard~\cite{ieee1997wireless}.}

\vspace{-3mm}
\subsection{Transmission Mode in Multi-Link Operation}\label{subsec:MLO}
\vspace{-.15mm}
Recalling that MLO adopts a distinct two-tier MAC architecture: the unified U-MAC serves as the common component for all links, while each link has its own independent L-MAC. The L-MAC performs channel access according to one of two transmission models: (i) simultaneous transmit and receive (STR) and (ii) non-simultaneous transmit and receive (NSTR). Specifically, STR enables independent link operations with asynchronous data transmission and reception, allowing each link to maintain its own channel access parameters. Namely, under STR, APs/STAs perform the backoff process independently on each link, and transmission attempts across different links are mutually independent. NSTR requires synchronized data transmissions across all available links by enforcing either end-time alignment or the defer-transmission mechanism~\cite{IEEE80211beD7}. In this paper, we focus on the STR mode, as it offers superior throughput performance~\cite{lopez2022multi} and is recommended as the default operational scheme in Wi-Fi 7.

\begin{table}[!t]
\centering
\footnotesize
\caption{Main Notations.}
\label{tab:notations}
\setlength{\tabcolsep}{6pt}
\begin{tabular}{cp{6.3cm}}
\toprule
\textbf{Notation} & \textbf{Description} \\
\midrule
$b_{i,k}^{(n,l)}$ & The stationary probability of the Markov chain for STA $n$ on link $l$ in backoff stage $i$ with backoff counter $k$.\\
$b_{n,l}(t)$ & The stochastic process representing the backoff counter of STA $n$ on link $l$ at time slot $t$.\\
$C_e$ & The cell state of LSTM cell   $e$.\\
$f_e$ & The output of the forget gate of LSTM cell   $e$.\\
$h_e$ & The hidden state output of LSTM cell   $e$.\\
$j_e$ & The updated input activation of LSTM cell   $e$.\\
$\mathcal{L}$ & The set of available links.\\
$\mathcal{N}$ & The set of STAs.\\
$\mathcal{T}$ & The set of time slots.\\
$B_e$ & The output gate control signal of LSTM cell   $e$.\\
$p_{n,l}$ & The probability that STA $n$ experiences a collision on link $l$.\\
$P_{l}^{tr}$ & The probability that at least one STA transmits on link $l$ in a randomly chosen time slot.\\
$P_{l}^{su}$ & The probability that a transmission on link $l$ is successful, conditioned on at least one transmission occurring.\\
$q_{n,l}(t)$ & The stochastic process representing the backoff stage of STA $n$ on link $l$ at time slot $t$.\\
$\tau_{n,l}$ & The probability that STA $n$ transmits a packet on link $l$ in a randomly chosen time slot.\\
$\widetilde{\mathcal{R}}(d)$ & The total network throughput achieved during decision step $d$.\\
$\beta_{n,l}(t)$ & The probability that a packet from STA $n$ is allocated to link $l$ at time slot $t$.\\
$w_{n,l}^0(t)$ & The ICW size when STA $n$ first attempts to access link $l$ at time slot $t$.\\
\bottomrule
\end{tabular}
\end{table}

\vspace{-3mm}
\subsection{Network Outline}\label{subsec:networkoutline}
\vspace{-.15mm} 
As shown in Fig.~\ref{fig:MLO MAC}(a), we consider an uplink (STA-to-AP) Wi-Fi 7 network with $L$ links, denoted by the set $\mathcal{L}=\{1,\ldots,L\}$, consisting of a single AP and $N$ STAs, denoted by the set $\mathcal{N}=\{1,\ldots,N\}$. All STAs operate with saturated traffic under the STR mode. Upon a data packet’s arrival at the MAC of STA $n$ at time slot $t$, the U-MAC assigns the packet from the buffer to a specific link according to the \emph{{traffic allocation policy}} $\beta_{n,l}(t)$, which denotes the probability that the packet is allocated to link $l$. Subsequently, at the L-MAC, each link performs channel access independently based on its assigned \emph{{ICW size}} $w^0_{n,l}(t)$. That is, when STA $n$ attempts to access link $l$ at first time, it uniformly selects a backoff counter from the interval $[0,w^0_{n,l}(t)-1]$. \emph{{As a result, jointly optimizing these two components, i.e., $\beta_{n,l}(t)$ at the U-MAC layer and $w^0_{n,l}(t)$ at the L-MAC layer, constitutes the central motivation of this paper.}} The major notations used in the paper are summarized in Table~\ref{tab:notations}.

\vspace{-3mm}
\subsection{Throughput Analysis via Bianchi Markov Chain}\label{subsec:analyticalframework}
\vspace{-.15mm} 

We extend the widely adopted Bianchi Markov model for single-link Wi-Fi~\cite{bianchi2002performance} to the MLO setting with $L$ links, deriving expressions that relate network throughput to both the traffic allocation policy and the ICW sizes. This extension is enabled by the observation that, under STR mode, the contention processes on different links can be decoupled because they operate over separate frequency bands, while the channel access procedure on each individual link still follows the standard CSMA/CA backoff mechanism assumed in Bianchi’s analysis. In constructing the Markov model, we make the following standard assumptions: (i) transmission failures occur solely due to collisions; and (ii) the number of STAs remains fixed.

\begin{figure}[t!]
\includegraphics[width=.45\textwidth]{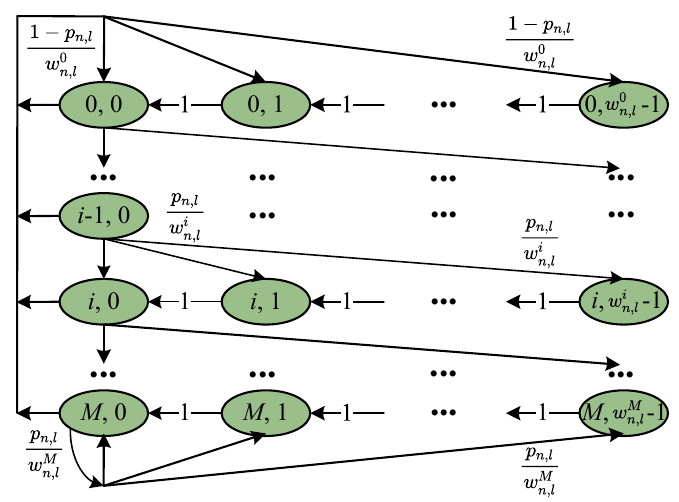}
\centering
\vspace{-1.5mm}
\caption{Markov chain model for the backoff stage $q_{n,l}(t)=i$ and backoff counter $b_{n,l}(t)=k$ of STA $n$ on link $l$.}
\label{fig:Markovchain}
\end{figure}

The analysis begins by modeling the behavior of STA $n$ on link $l$ using a Markov chain, from which we derive the stationary transmission probability $\tau_{n,l}$, i.e., the probability that STA $n$ transmits a packet on link $l$ in a randomly chosen time slot. Specifically, let $b_{n,l}(t)$ denote the stochastic process representing the backoff counter of STA $n$ on link $l$ at time slot $t$,\footnote{In this paper, a time slot $t$ refers to the moment when the backoff counter is decremented. Note that this does not correspond directly to the fixed Wi-Fi time slot, because the backoff countdown pauses whenever the channel is busy (e.g., during packet transmissions). Therefore, throughout the paper, time slot $t$ may represent either the fixed Wi-Fi time slot or the variable interval between two consecutive backoff counter decrements.} and let $q_{n,l}(t)$ denote the stochastic process representing the backoff stage $(0,\ldots,M)$, i.e., the number of unsuccessful transmission attempts. Assuming that, at each transmission attempt and regardless of the number of retransmissions, STA $n$ on link $l$ experiences a collision with probability $p_{n,l}$, the bidimensional process $\{q_{n,l}(t), b_{n,l}(t)\}$ can be modeled as the discrete-time Markov chain shown in Fig.~\ref{fig:Markovchain}. In this Markov chain, the one-step transition probabilities are given by\footnote{For simplicity, we adopt the shorthand notation $P\{i_1,k_1 \mid i_0,k_0 \}=P\{q_{n,l}(t+1)=i_1,b_{n,l}(t+1)=k_1 \mid q_{n,l}(t)=i_0,b_{n,l}(t)=k_0 \}$.}
\vspace{-.5mm}
\begin{equation}\label{eq:transition}
\begin{cases}
P\{ i,k \mid i,k+1 \} = 1, 
&\hspace{-2mm} k \in [0, w_{n,l}^i - 2],\; i \in [0,M],\\[3pt]
P\{ 0,k \mid i,0 \} = \dfrac{1-p_{n,l}}{w_{n,l}^0}, 
&\hspace{-2mm} k \in [0, w_{n,l}^0 - 1],\; i \in [0,M],\\[8pt]
P\{ i,k \mid i-1,0 \} = \dfrac{p_{n,l}}{w_{n,l}^i}, 
&\hspace{-2mm} k \in [0, w_{n,l}^i - 1],\; i \in [1,M],\\[7pt]
P\{ M,k \mid M,0 \} = \dfrac{p_{n,l}}{w_{n,l}^M},
&\hspace{-2mm} k \in [0, w_{n,l}^M - 1]. 
\end{cases}
\end{equation} 
Here, the first equation captures the decrement of the backoff counter at the beginning of each time slot. The second equation states that after a successful transmission, a new packet enters backoff stage 0 with its backoff counter uniformly selected from $[0, w_{n,l}^0 - 1]$. The third and fourth equations model unsuccessful transmissions: following a failure at backoff stage $i-1$, the STA moves to backoff stage $i$ and draws a new backoff counter from $[0, w_{n,l}^i - 1]$, whereas once the maximum backoff stage $M$ is reached, the backoff stage is no longer increased.


Let $b^{(n,l)}_{i,k}= \lim_{t \rightarrow \infty}P\{q_{n,l}(t)=i,b_{n,l}(t)=k \}$ denote the stationary distribution of the Markov chain for STA $n$ on link $l$. Owing to the regularity properties of the chain, we can obtain~\cite{bianchi2002performance}
\vspace{-.5mm}
\begin{equation}\label{eq:stationary}
    b^{(n,l)}_{i,k} = \frac{w_{n,l}^i - k}{w_{n,l}^i} b^{(n,l)}_{i,0}, \quad i\in[0,M],\; k\in[0,w_{n,l}^i-1].
\end{equation}

Since the stationary probabilities of the Markov chain must sum to one, applying this normalization to~\eqref{eq:stationary} yields
\vspace{-.5mm}
\begin{align}\label{eq:sumstationary}
\sum_{i=0}^{M} \sum_{k=0}^{w_{n,l}^i-1}b^{(n,l)}_{i,k}= \sum_{i=0}^{M}b^{(n,l)}_{i,0}\sum_{k=0}^{w_{n,l}^i-1} \frac{w_{n,l}^i-k}{w_{n,l}^i}=1.
\end{align}
Then, using the facts that $\sum_{i=0}^{M} b^{(n,l)}_{i,0}(1-p_{n,l}) = b^{(n,l)}_{0,0}$ and $w_{n,l}^i = 2^i w_{n,l}^0$ under the exponential backoff scheme of CSMA/CA, we can derive $b^{(n,l)}_{0,0}$ from~\eqref{eq:sumstationary} as follows:
\vspace{-.5mm}
\begin{equation}\label{eq:b0}
  \hspace{-1mm}  b_{0,0}^{(n,l)} \!\!=\! \frac{2(1-2p_{n,l})(1-p_{n,l})}{(1-2p_{n,l})(w_{n,l}^0+1)\! +\! p_{n,l} w_{n,l}^0 \bigl(1 - (2p_{n,l})^M \bigr)}.
\end{equation}

Recalling that the traffic allocation policy $\beta_{n,l}$ denotes the probability that a data packet from STA $n$ is allocated to link $l$, we can now express the transmission probability $\tau_{n,l}$—the probability that STA $n$ transmits on link $l$ in a randomly selected time slot. Since a transmission occurs when the traffic is allocated to link $l$ and the backoff counter reaches zero, regardless of the backoff stage, we have
\vspace{-.5mm}
\begin{align}\label{eq:transmitprob}
    &\tau_{n,l} = \beta_{n,l}\sum_{i=0}^{M} b_{i,0}^{(n,l)} = \beta_{n,l}\frac{b_{0,0}^{(n,l)}}{1-p_{n,l}} \nonumber\\
    &= \frac{2(1-2p_{n,l})\beta_{n,l}}{(1-2p_{n,l})(w_{n,l}^0+1) + p_{n,l} w_{n,l}^0\bigl(1 - (2p_{n,l})^{M}\bigr)}.
\end{align}

Meanwhile, the collision probability $p_{n,l}$ experienced by STA $n$ on link $l$ equals the probability that at least one of the remaining $N-1$ STAs transmits on link $l$. This yields
\vspace{-.5mm}
\begin{align}\label{eq:collisionprob}
    p_{n, l} = 1 - \prod_{\substack{m=1, m \ne n}}^{N} (1 - \tau_{m, l}).
\end{align}

Let $P_{l}^{tr}=1 - \prod_{\substack{n=1}}^{N} (1 - \tau_{n, l})$ denote the probability that at least one STA transmits on link $l$. Let $P_{l}^{su}=\frac{\sum_{n=1}^{N}\tau_{n,l}\prod_{\substack{m=1, m\neq n}}^{N}(1-\tau_{m,l})}{P_{l}^{tr}}$ be the probability that a transmission on link $l$ is successful (i.e., exactly one STA transmits, conditioned on at least one transmission occurring). Using these, the normalized throughput of link $l$ -- defined as the fraction of time during which the channel is used to successfully transmit payload bits -- can be expressed as
\vspace{-.5mm}
\begin{equation}\label{eq:throughputperlink}
\mathcal{R}_{l}=\frac{  P^{tr}_l P_{l}^{su}E\left[Y_{l}\right]}{\left(1-P^{tr}_l\right) \sigma+P^{tr}_l P_{l}^{su} T_{l}^{su}+P^{tr}_l\left(1-P_{l}^{su}\right) T_{l}^{co}},
\end{equation}
where $E\left[Y_{l}\right]$ denotes the average packet payload size transmitted on link $l$, $\sigma$ is the duration of an empty slot, $T_{l}^{su}$ is the average time that link $l$ is sensed busy due to a successful transmission, and $T_{l}^{co}$ is the average time that link $l$ is sensed busy due to a collision.\footnote{The durations of $T_{l}^{su}$ and $T_{l}^{co}$ are determined by the PHY/MAC–layer timing parameters such as SIFS, AIFS, ACK transmission time, and the PHY header duration. These parameters are fixed prior to the simulation and remain constant throughout the evaluation.}  

Under the STR mode, the throughputs of different links are independent since they operate on separate frequency bands. Therefore, the total network throughput is obtained by summing the throughputs of all links, which can be expressed as
\vspace{-.5mm}
\begin{equation}\label{eq:networkthroughput}
\widetilde{\mathcal{R}}=\sum_{l=1}^L\mathcal{R}_{l}.
\end{equation}
{Based on~\eqref{eq:transition}–\eqref{eq:networkthroughput}, the total network throughput $\widetilde{\mathcal{R}}$ is expressed as a function of the traffic allocation policy $\beta_{n,l}$ at the U-MAC and the ICW $w_{n,l}^0$ at the L-MAC. \emph{\textbf{This clearly demonstrates the necessity of cross-layer optimization to fully unlock the performance potential of Wi-Fi 7 MLO.}}}

\vspace{-3mm}
\subsection{Problem Formulation}\label{subsec:probformulate}
\vspace{-.15mm}

The goal of this paper is to maximize the throughput in~\eqref{eq:networkthroughput} -- consistent with the primary performance objective of Wi-Fi 7 (i.e., extremely high throughput) -- by jointly optimizing the traffic allocation policy $\beta_{n,l}(t)$ and the ICW size $w^0_{n,l}(t)$, i.e.,
\vspace{-1.5mm}
\begin{align}
      &\max_{\{\bm{\beta},\mathbf{w}\}} \sum_{d=1}^D\widetilde{\mathcal{R}}(d)\hspace{-40mm}\label{eq:problem} \\
     &\text{s.t.} \quad \forall n\in \mathcal{N},\ l \in \mathcal{L}, \ t\in \mathcal{T}, \nonumber\\
     &\mathcal{C}1: \  \sum_{l=1}^L \beta_{n,l}(t)=1,  \nonumber \\
     &\mathcal{C}2:  \ w_{min}\leq w^0_{n,l}(t) \leq w_{max}, \nonumber 
\end{align}
where $\bm{\beta}=\{\beta_{n,l}(t)\}_{n\in \mathcal{N},l\in\mathcal{L},t\in \mathcal{T}}$ denotes the traffic allocation policy of different STAs across links over time, $\mathbf{w}=\{w^0_{n,l}(t)\}_{n\in \mathcal{N},l\in\mathcal{L},t\in \mathcal{T}}$ represents the ICW sizes of different STAs on different links over time, and $d=1,2,\ldots,D$ denotes the decision steps at which the traffic allocation policy and ICW sizes are updated (as detailed in Sec.~\ref{subsec:MDP}).\footnote{Each decision step $d$ corresponds to one period of the real-world CSMA/CA procedure, which is set to 20 ms in our experiments and comprises multiple time slots $t$. As a result, the stationary transmission probability $\tau_{n,l}$ remains constant within each decision step, and the network throughput $\widetilde{\mathcal{R}}(d)$ can be obtained by evaluating the bidimensional process ${q_{n,l}(t), b_{n,l}(t)}$ according to~\eqref{eq:transition}–\eqref{eq:networkthroughput}.}  

Constraint $\mathcal{C}1$ ensures that all outgoing traffic of STA $n$ is fully allocated across the available links without duplication or omission, while constraint $\mathcal{C}2$ guarantees that the ICW adaptation remains compliant with protocol specifications, where $w_{min}$ and $w_{max}$ denote the minimum and maximum CW sizes, respectively.

\vspace{-.5mm}
\begin{remark}\label{rem:NPhard} As shown in~\eqref{eq:transmitprob}, the transmission probability $\tau_{n,l}$ depends jointly on the traffic allocation policy $\beta_{n,l}$ at the U-MAC layer and the ICW size $w^0_{n,l}$ at the L-MAC layer, while the collision probability $p_{n,l}$ recursively depends on $\tau_{n,l}$ as expressed in~\eqref{eq:collisionprob}. In addition, network dynamics evolve over time due to varying traffic loads and fluctuating contention levels, creating temporal dependencies among decisions. {Consequently, the problem in~\eqref{eq:problem} becomes a non-convex, nonlinear program with recursive cross-layer coupling and strong temporal correlations, making direct optimization computationally prohibitive. This complexity motivates the use of a DRL-based solution (as detailed in Sec.~\ref{sec:LSTMSAC}). }
\end{remark}
\vspace{-.5mm}

\vspace{-3mm}
\section{Basic Idea of Long Short-Term Memory} \label{sec:LSTM}
\vspace{-.15mm}
Before introducing our LSTM-SAC algorithm, we first explain the motivation for integrating long short-term memory (LSTM) with deep reinforcement learning (DRL). We then detail the components of each LSTM cell.

\vspace{-3mm}
\subsection{Motivation of Adopting LSTM}  \label{subsec:LSTMmotivation} 
\vspace{-.15mm}
{Applying conventional DRL to Wi-Fi 7 MLO faces two major challenges. \emph{{First, the decision-making process is inherently partially observable.}} In practical Wi-Fi 7 networks, the DRL agent cannot directly observe the exact backoff states of competing STAs or instantaneous collision events on each link. Consequently, the problem is more accurately modeled as a partially observable Markov decision process (POMDP)~\cite{wydmanski2021contention}, where the underlying network state must be inferred from noisy and incomplete observations. \emph{{Second, the system exhibits non-Markovian behavior.}} As shown in Fig.~\ref{fig:nonMarkovian}, the performance of an action depends not only on the current state but also on actions taken in subsequent states. That is, the network throughput achieved by a given traffic allocation policy and ICW size selected upon the arrival of the current flow is not solely determined by this decision. It is further influenced by the actions taken for subsequent flows, as later arrivals modify channel congestion and the resulting signal-to-noise ratio (SNR). Consequently, these two factors significantly degrade the effectiveness and training stability of conventional DRL methods.}

The LSTM~\cite{hochreiter1997long,chen2024information,cui2022qos}, a variant of recurrent neural networks (RNNs) that incorporates gated mechanisms (as detailed in Sec.~\ref{subsec:LSTM}), is capable of capturing temporal dependencies in sequential data. When integrated into DRL, LSTM enables the agent to encode a history of past observations into a compact latent representation, thereby approximating the underlying hidden system state. This capability allows the DRL agent to recover temporal correlations induced by non-Markovian dynamics and to infer unobservable network conditions under partial observability, leading to more stable training and improved decision quality in Wi-Fi~7 MLO networks.

\begin{figure}[t!]
\includegraphics[width=.46\textwidth]{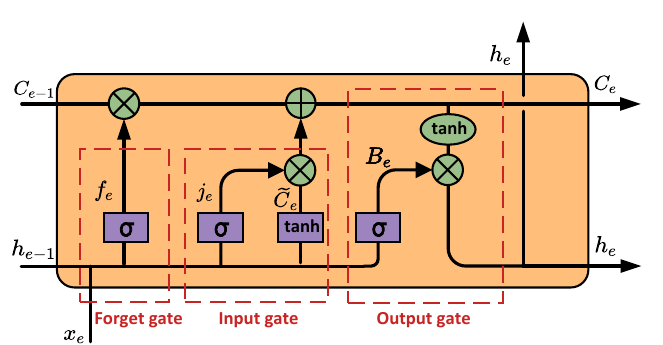}
\centering
\vspace{-1.5mm}
\caption{Internal structure of an LSTM cell, containing the forget gate, input gate, and output gate.}
\label{fig:LSTMcell}
\end{figure}
\begin{figure}[t!]
\includegraphics[width=.45\textwidth]{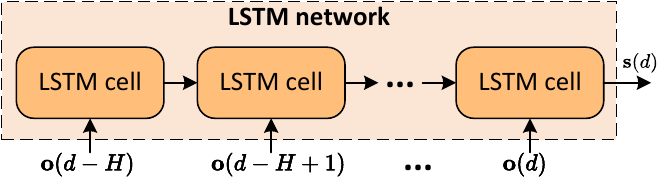}
\centering
\vspace{-1.5mm}
\caption{LSTM-based partial historical information state representation layer network.}
\label{fig:LSTMnetwork}
\end{figure}

\vspace{-3mm}
\subsection{Preliminaries of LSTM}  \label{subsec:LSTM}
\vspace{-.15mm}

An LSTM cell consists of two states, namely the \emph{cell state} and the \emph{hidden state}, as well as three gates: the \emph{forget gate}, \emph{input gate}, and \emph{output gate}, as illustrated in Fig.~\ref{fig:LSTMcell}. We denote the outputs of the forget gate, cell state, and hidden state at cell $e$ by $f_e$, $C_e$, and $h_e$, respectively. The forget gate implements the forgetting mechanism by determining which information from the previous cell state should be discarded, which can be expressed as
\vspace{-.5mm}
\begin{align}\label{eq:forget}
    f_{e} = \text{sigmod}\big(W^x_{f}x_e+W^h_{f}h_{e-1} +\xi_f\big),
\end{align}
where $x_e$ denotes the input to the current cell.

The input gate processes the current input and determines what information should be stored into the cell state. This operation consists of two parts: a sigmoid layer that computes the updated value $j_e$, and a $\tanh$ layer that generates the candidate cell state $\widetilde{C}_e$, given by
\vspace{-.5mm}
\begin{align}\label{eq:input}
    j_{e} &= \text{sigmod}\big(W^x_{j}x_e+W^h_{j}h_{e-1} +\xi_j\big),\\
    \widetilde{C}_e &= \tanh \big(W^x_{C}x_e+W^h_{C}h_{e-1} +\xi_C \big).
\end{align}
The cell state is then updated by combining these two components:
\vspace{-.5mm}
\begin{align}\label{eq:cellstate}
    C_e=f_e\odot C_{e-1}+j_{e}\odot\widetilde{C}_e,
\end{align}
where $f_e \odot C_{e-1}$ selectively retains past information, and $j_e \odot \widetilde{C}_e$ selectively incorporates new information. Here, $\odot$ denotes element-wise multiplication.

The output gate determines the output of the current LSTM cell, enabling it to selectively retain relevant information while suppressing irrelevant features. The control signal of the output gate, denoted by $B_e$, is first computed using a sigmoid function, and the current hidden state is then obtained as follows:
\vspace{-.5mm}
\begin{align}\label{eq:output}
    B_e &= \text{sigmod}\big(W^x_{B}x_e+W^h_{B}h_{e-1} +\xi_B\big),\\
    h_e &= B_e\odot \tanh(C_e).
\end{align}
Here, $\{W^x_{f}, W^h_{f}, W^x_{j}, W^h_{j}, W^x_{C}, W^h_{C}, W^x_{B}, W^h_{B}\}$ and $\{\xi_f, \xi_j, \xi_C, \xi_B\}$ denote the weight matrices and bias vectors. These parameters constitute the trainable components of the LSTM network and can be uniformly expressed as $\eta = \big\{W^x_{f},W^h_{f},W^x_{j},W^h_{j},W^x_{C},W^h_{C}, W^x_{B},W^h_{B},\xi_f,\xi_j,\xi_C,\xi_B\big\}$.

Finally, as shown in Fig.~\ref{fig:LSTMnetwork}, we propose an LSTM-based historical information representation layer composed of multiple LSTM cells and integrated with the conventional DRL framework. Specifically, the LSTM network takes $H$ sequential historical observations as input and outputs an aggregated latent state, which serves as the state representation for DRL decision-making (as detailed in Sec.~\ref{subsec:MDP}).

\vspace{-3mm}
\section{LSTM-based Soft Actor–Critic Algorithm}\label{sec:LSTMSAC}
\vspace{-.15mm}
In this section, we first present the motivation of adopting the soft actor-critic (SAC) algorithm. We then define the elements of the Markov decision process (MDP), followed by an overview of the LSTM-SAC algorithm’s architecture. Finally, we provide a comprehensive analysis of its computational complexity.

\vspace{-3mm}
\subsection{Motivation for Adopting the SAC Algorithm}
\vspace{-.15mm}

The primary motivation for adopting a DRL-based solution arises from the fact that the formulated problem in~\eqref{eq:problem} is highly non-convex and nonlinear (as discussed in Remark~\ref{rem:NPhard}), making direct optimization computationally prohibitive. Existing analytical approaches~\cite{gao2025latency, dai2012unified,park2024adaptive,11373758,hou2026energy} also face significant scalability challenges (as detailed in Sec.~\ref{subsec:challenges}).

Among various DRL algorithms, we select SAC~\cite{haarnoja2018soft} due to its superior stability, efficiency, and exploration capability. SAC is built on the maximum-entropy reinforcement learning framework, where entropy reflects the randomness of the agent’s policy~\cite{haarnoja2018soft}. Unlike traditional DRL methods such as DQN~\cite{mnih2013playing} and DDPG~\cite{lillicrap2015continuous,liu2025lyapunov}, which focus solely on maximizing cumulative rewards and often suffer from insufficient exploration and unstable training, SAC simultaneously maximizes the expected return and policy entropy. This entropy-regularized objective encourages persistent exploration, significantly enhances learning stability, and mitigates the risk of converging to suboptimal solutions—particularly important in high-dimensional and dynamic Wi-Fi 7 MLO networks.

\vspace{-3mm}
\subsection{MDP Elements in the LSTM-SAC Algorithm} \label{subsec:MDP}
\vspace{-.15mm}
We deploy a DRL agent at the AP, where the agent learns to select an \emph{action} based on the observed system \emph{state} under a trained policy by interacting with the environment and receiving a corresponding \emph{reward}:

\begin{enumerate}[leftmargin=4mm]
\item \textbf{\textit{State Space:}} At decision step $d$, the DRL agent obtains the network information through observation $\mathbf{o}(d)=\{\mathbf{g}(d),\mathbf{u}(d))\}$, where $\mathbf{g}(d)=\{g_{n,l}(d)\}_{n\in\mathcal{N},l\in\mathcal{L}}$ denotes the SNR of each STA–link pair, and $\mathbf{u}(d)=\{u_{n,l}(d)\}_{n\in\mathcal{N},l\in\mathcal{L}}$ represents the corresponding channel busy time. We then select $H$ sequential historical observations as the input to the LSTM network and use its output as the state representation. Accordingly, the state is defined as:
\vspace{-.5mm}
\begin{align}\label{eq:state}
  \hspace{-2mm}  \mathbf{s}(d)=\{LSTM(\mathbf{o}(d-H),\ldots, \mathbf{o}(d))\}, \ d \geq H.
\end{align}
As shown in Fig.~\ref{fig:timeline}, as the number of decision steps increases, the oldest observation in the history window is removed and the newest observation is appended. In addition, each decision step $d$ corresponds to one period of the real-world CSMA/CA procedure, which is set to 20 ms in our experiments. 


\begin{figure}[!t]
\includegraphics[width=.45\textwidth]{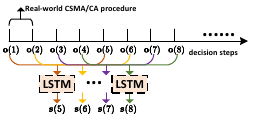}
\centering
\vspace{-1.5mm}
\caption{The relationship between the observation $\mathbf{o}(d)$ and state $\mathbf{s}(d)$ when the history window is set to $H=5$.}
\label{fig:timeline}
\end{figure}

\begin{figure}[!t]
\includegraphics[width=.47\textwidth]{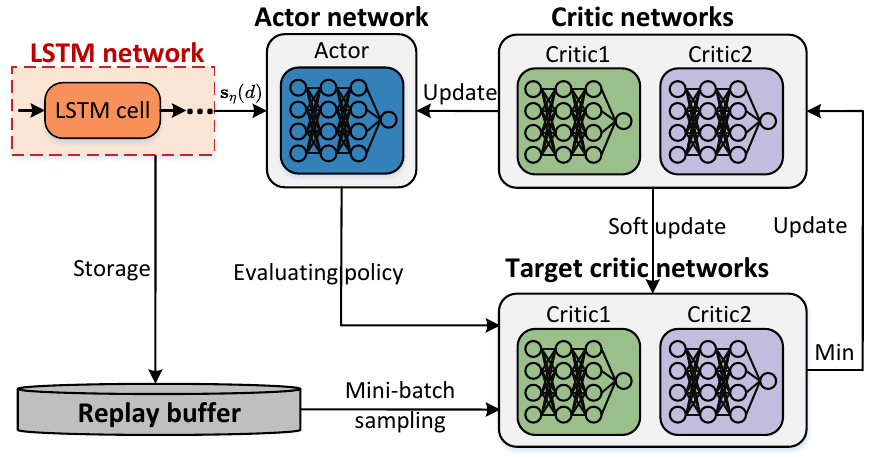}
\centering
\vspace{-1.5mm}
\caption{The overall architecture of the LSTM-SAC algorithm.}
\label{fig:LSTM-SAC}
\end{figure}

\item \textbf{\textit{Action Space:}} Based on state $\mathbf{s}(d)$, the DRL agent selects an action at each decision step $d$ to interact with the environment. The action space contains $2NL$ elements, and is expressed as follows:
\vspace{-.5mm}
\begin{align}\label{eq:action}
    \mathbf{a}(d)=\{\bm{\beta}(d),\mathbf{w}(d)\},
\end{align}
where $\bm{\beta}(d)=\{\beta_{n,l}(d)\}_{n\in \mathcal{N},l\in\mathcal{L}}$ denotes the traffic allocation vector, specifying for each STA the probability of assigning a packet to each link at decision step $d$, and $\mathbf{w}(d)=\{w^0_{n,l}(d)\}_{n\in \mathcal{N}, l\in\mathcal{L}}$ represents the ICW sizes for different STAs on different links at decision step $d$. Specifically, the initial traffic allocation policy $\widetilde{\bm{\beta}}(d)$ is first normalized to the range $[0,1]$. To satisfy constraint $\mathcal{C}1$, we enforce $\beta_{n,l}(d)=\frac{\widetilde{\beta_{n,l}}(d)}{\sum_{l \in \mathcal{L}}\widetilde{\beta_{n,l}}(d)}$. Similarly, the ICW size $\widetilde{\mathbf{w}}(d)$ is normalized to the range $[0,6]$. To satisfy constraint $\mathcal{C}2$, we map it to $w^0_{n,l}(d)=\lfloor2^{\widetilde{w^0_{n,l}}(d)+4}\rfloor$, which ensures that $w^0_{n,l}(d)$ falls within the standard CW range of 16 to 1024. Note that within each decision step $d$, the updated traffic allocation policy $\bm{\beta}(d)$ and ICW sizes $\mathbf{w}(d)$ remain fixed throughout the 20 ms CSMA/CA period, until the next decision step $d+1$ is triggered. 

\item \textbf{\textit{Reward Function:}} After executing action $\mathbf{a}(d)$ based on the state $\mathbf{s}(d)$, the environment returns a reward $r(d)$. As detailed in Sec.~\ref{subsec:analyticalframework}, we adopt the system objective in \eqref{eq:networkthroughput} as the reward, which can be given by
\vspace{-.5mm}
\begin{align}\label{eq:LSTM-SACreward}
    r(d)=\widetilde{\mathcal{R}}(d),
\end{align}
with the goal of maximizing network throughput, serving as the primary performance objective of Wi-Fi 7 (i.e., extremely high throughput).

\end{enumerate}

\subsection{Architecture of the LSTM-SAC Algorithm}  \label{subsec:architecture}
The architecture of LSTM-SAC is illustrated in Fig.~\ref{fig:LSTM-SAC}. It comprises an LSTM network, an actor network, two critic networks, two target critic networks, and a replay buffer, as detailed below.

\begin{enumerate}[leftmargin=4mm]
\item \textbf{\textit{LSTM Network:}} The LSTM network, parameterized by $\eta=\{W^x_{f}, W^h_{f}, W^x_{j}, W^h_{j}, W^x_{C}, W^h_{C}, W^x_{B}, W^h_{B}, \xi_f, \xi_j, \xi_C, \xi_B\}$, takes $H$ sequential historical observations as input and outputs a corresponding state representation $\mathbf{s}_{\eta}(d)$. For traffic allocation and ICW size decisions, the state representation $\mathbf{s}_{\eta}(d)$ produced by the LSTM is then fed into the SAC algorithm and combined with reward signals to guide policy optimization.

\item \textbf{\textit{Actor Network:}} The actor network $\pi_{\phi}$, parameterized by $\phi$, takes the state $\mathbf{s}_{\eta}(d)$ as input and outputs the corresponding action $\mathbf{a}(d)$. Unlike the standard reinforcement learning objective, which focuses solely on maximizing the expected cumulative reward, LSTM-SAC further incorporates an entropy term. This enables the optimal policy to maximize its entropy, thereby maintaining sufficient exploration~\cite{haarnoja2018soft}:
\vspace{-.5mm}
\begin{align}\label{eq:optimalpolicy}
    \pi^*_{\phi}=\arg \max_{\pi_{\phi}}\sum_{d}\mathbb{E}[r(d)+\varepsilon \mathcal{H}(\pi_{\phi}(\cdot | \mathbf{s}_{\eta}(d)))],
\end{align}
where $\mathcal{H}(\pi_{\phi}(\cdot | \mathbf{s}_{\eta}(d)))=\mathbb{E}[-\log\pi_{\phi}(\mathbf{a}(d)|\mathbf{s}_{\eta}(d))]$ denotes the entropy that measures the randomness of policy $\pi_{\phi}$, and $\varepsilon$ is the temperature parameter that controls the trade-off between the entropy term and the reward.

\item \textbf{\textit{Replay Buffer:}} During training, at each decision step $d$, the transition tuple $[(\mathbf{o}(d-H),\ldots, \mathbf{o}(d)), \mathbf{a}(d), r(d), (\mathbf{o}(d-H+1),\ldots, \mathbf{o}(d+1))]$ is stored in the replay buffer $\mathcal{F}$. Notably, we store the $H$ sequential historical observations $(\mathbf{o}(d-H),\ldots,\mathbf{o}(d))$ rather than the encoded state $\mathbf{s}_{\eta}(d)$. This design ensures that the LSTM network can be trained jointly with SAC in an end-to-end manner, without breaking the computational graph during gradient backpropagation. The replay buffer then serves as an experience memory, allowing the DRL agent to decorrelate samples via random mini-batch sampling, which improves sample efficiency and stabilizes the training process.

\item \textbf{\textit{Two Critic Networks:}} Two critic networks, $Q_{\theta_1}$ and $Q_{\theta_2}$, parameterized by $\theta_1$ and $\theta_2$, respectively, take the state $\mathbf{s}_{\eta}(d)$ and action $\mathbf{a}(d)$ as inputs and produce the corresponding Q-values $Q_{\theta_m}(\mathbf{s}_{\eta}(d),\mathbf{a}(d)))$~\cite{haarnoja2018soft}:
\vspace{-.5mm}
\begin{align}\label{eq:softactionvalue}
    &Q_{\theta_m}(\mathbf{s}_{\eta}(d),\mathbf{a}(d))=r(d)+ \gamma\mathbb{E}[Q_{\theta_m}(\mathbf{s}_{\eta}(d+1),\mathbf{a}(d))) \nonumber\\
    &-\varepsilon\log\pi_{\phi}(\mathbf{a}(d+1)|\mathbf{s}_{\eta}(d+1))], \ m\in \{1,2\}.
\end{align}
Specifically, $\gamma$ denotes the reward discount factor, and these Q-values quantify the expected quality of the state–action pair by incorporating both reward and entropy, where a higher value reflects not only a greater expected return but also a higher potential for exploration. Moreover, during policy improvement, the smaller of the two Q-values is used as the target Q-value, which helps reduce overestimation bias and stabilize training.

\begin{algorithm} [!t]
  \footnotesize
  \SetAlgoLined
  \SetKwData{Left}{left}\SetKwData{This}{this}\SetKwData{Up}{up}
  \SetKwFunction{Union}{Union}\SetKwFunction{FindCompress}{FindCompress}
  \SetKwInOut{Input}{input}\SetKwInOut{Output}{output}
  \textbf{Input:} $\eta$, $\phi$, $\theta_1$, $\theta_2$, $\bar{\theta}_1 \leftarrow \theta_1$, $\bar{\theta}_2 \leftarrow \theta_2$, $\mathcal{F}\leftarrow \emptyset$.

  \textbf{Output:} The optimal traffic allocation policy and initial CW size.
  \BlankLine
  
  \For{$episode=1$ \KwTo $E$}{
    \For{$d=1$ \KwTo $H$}{
        Reset the environment, randomly select an action $\mathbf{a}(d)$, receive the reward $r(d)$ according to~\eqref{eq:LSTM-SACreward}, and obtain the next observation $\mathbf{o}(d+1)$.
    }
    \For{$d=H$ \KwTo $D$}{ 
        The $H$ sequential historical observations $\{\mathbf{o}(d-H),\ldots,\mathbf{o}(d)\}$ are fed into the LSTM network to generate the state $\mathbf{s}_{\eta}(d)$.

        Input the state $\mathbf{s}_{\eta}(d)$ into the actor network $\pi_{\phi}$ to produce the corresponding action $\mathbf{a}(d)$.
                
        Receive the reward $r(d)$ according to~\eqref{eq:LSTM-SACreward} and obtain the next observation $\mathbf{o}(d+1)$.

        Store the transition tuple $[(\mathbf{o}(d-H),\ldots, \mathbf{o}(d)), \mathbf{a}(d), r(d), (\mathbf{o}(d-H+1),\ldots, \mathbf{o}(d+1))]$ into $\mathcal{F}$.
        
        Delete the first observation in the history window and append $\mathbf{o}(d+1)$ to the end of the history window.

        The $H$ sequential historical observations $\{\mathbf{o}(d-H+1),\ldots,\mathbf{o}(d+1)\}$ are fed into the LSTM network to generate the state $\mathbf{s}_{\eta}(d+1)$.
        
        Randomly sample a mini-batch of transitions from the replay buffer. Update the parameters $\theta_1$ and $\theta_2$ of the two critic networks, together with the LSTM parameters $\eta$, by minimizing~\eqref{eq:LSTM-SACcritic}. Update the actor network parameters $\phi$ by minimizing~\eqref{eq:LSTM-SACactor}, and adaptively update the entropy coefficient $\varepsilon$ by minimizing~\eqref{eq:entropyloss}. Finally, update the parameters $\bar{\theta}_1$ and $\bar{\theta}_2$ of the target critic networks using~\eqref{eq:LSTM-SACtarget}.
        }
    }
  \caption{LSTM-SAC Algorithm.}\label{algo:LSTM-SAC}
\end{algorithm}\DecMargin{1em}

\item \textbf{\textit{Policy Improvement:}} After sufficient exploration, a mini-batch of samples is randomly drawn from the replay buffer $\mathcal{F}$ to update the critic, LSTM, and actor networks. Since the LSTM is responsible for extracting state representations that are critical for value estimation, its parameters should be trained using a stable and informative learning signal. Compared to the high-variance policy gradient from the actor, the value-based learning signal from the critic provides more reliable gradients for optimizing the LSTM. Therefore, the parameters of the two critic networks and the LSTM network are updated jointly using the same optimization objective. 

Specifically, the parameters $\theta_1$ and $\theta_2$ of two critic networks, together with the LSTM parameters $\eta$, are trained to minimize the soft Bellman residual \cite{haarnoja2018soft}:
\vspace{-.5mm}
\begin{align}\label{eq:LSTM-SACcritic}
    &J_{Q}(\theta_m,\eta)=\mathbb{E}_{\mathcal{F}}\Big[\frac{1}{2}(Q_{\theta_m}(\mathbf{s}_{\eta}(d),\mathbf{a}(d)) \nonumber\\ 
    &-\min_{m \in \{1,2\}} Q_{\bar{\theta}_m}(\mathbf{s}_{\eta}(d),\mathbf{a}(d)))^2\Big], \ m \in \{1,2\},
\end{align}
where $\min_{m \in {1,2}} Q_{\bar{\theta}_m}(\mathbf{s}_{\eta}(d),\mathbf{a}(d))$ denotes the soft Bellman target, defined as
\vspace{-.5mm}
\begin{align}\label{eq:Bellmantarget}
    &r(d)+\gamma \mathbb{E}\Big[
    \min_{m \in \{1,2\}} Q_{\bar{\theta}_m}(\mathbf{s}_{\eta}(d+1),\mathbf{a}(d)) 
    \nonumber\\
    &- \varepsilon\log\pi_{\phi}(\mathbf{a}(d+1)|\mathbf{s}_{\eta}(d+1))\Big],
\end{align}
where $Q_{\bar{\theta}_m}$ denotes the target critic networks, which take the next state $\mathbf{s}_{\eta}(d+1)$ and action $\mathbf{a}(d+1)$ as inputs and output the corresponding Q-values. The entropy term encourages policy stochasticity and improves exploration. 

The parameters $\phi$ of the actor network are then updated by minimizing the expected Kullback–Leibler divergence between the policy and the exponentiated Q-function \cite{haarnoja2018soft}:
\vspace{-.5mm}
\begin{align}\label{eq:LSTM-SACactor}
   \hspace{-3mm} J_{\pi}(\phi) = \mathbb{E}_\mathcal{F} [ 
    \mathbb{E} 
    [ \varepsilon \log \pi_{\phi}(\mathbf{a}(d)|\mathbf{s}_{\eta}(d)) \nonumber\\
    - Q_{\theta_1}(\mathbf{s}_{\eta}(d), \mathbf{a}(d)) ]].
\end{align}

The entropy coefficient $\varepsilon$ plays a critical role in balancing exploration and exploitation in SAC. A smaller $\varepsilon$ reduces the algorithm to a conventional actor–critic method that prioritizes reward maximization, whereas a larger $\varepsilon$ overly emphasizes entropy and may degrade performance. To automatically balance this trade-off, $\varepsilon$ is adaptively adjusted during training by minimizing the entropy loss:
\vspace{-.5mm}
\begin{align}\label{eq:entropyloss}
  J(\varepsilon) = \varepsilon\mathbb{E}_{\mathcal{F}}\big[\mathcal{H}\big(\pi_{\phi}(\cdot | \mathbf{s}_{\eta}(d))-\bar{\mathcal{H}}\big)\big],
\end{align}
where $\bar{\mathcal{H}}$ denotes the target entropy. 

Finally, to stabilize learning, the parameters of the target critic networks are updated using soft updates:
\vspace{-.5mm}
\begin{align}\label{eq:LSTM-SACtarget}
    \bar{\theta}_m\leftarrow \rho \theta_m+(1-\rho)\bar{\theta}_m, \ m \in \{1,2\}, 
\end{align}
where $\rho \in (0,1]$ is the target update rate. This gradual update mechanism ensures smooth evolution of the target Q-values and improves training stability.

\end{enumerate}

\vspace{-3mm}
\subsection{LSTM-SAC Algorithm and Complexity Analysis}
\vspace{-.15mm}
Algorithm~\ref{algo:LSTM-SAC} provides the pseudocode of the proposed LSTM-SAC algorithm. The computational complexity of LSTM-SAC mainly stems from two aspects: training complexity and execution complexity. Note that since the training of learning-based methods can be carried out in a cloud data center with abundant computational resources~\cite{liu2024dnn,11223083}, our analysis of the LSTM-SAC algorithm’s computational complexity primarily focuses on the execution phase.
 
During execution, only the LSTM network and the SAC actor network are required, and the computational cost corresponds to a single forward pass through these networks. According to~\cite{chen2024information}, the computational complexity of the LSTM network is $\mathcal{O}(L_{LSTM} M_{LSTM}^{2} C_{LSTM}^{2})$,\footnote{In this subsection, we slightly abuse the notation for simplicity.} where $L_{LSTM}$ denotes the number of hidden layers, $M_{LSTM}^{2}$ the number of memory blocks, and $C_{LSTM}^{2}$ the capacity of the memory cells in each block. In addition, for the SAC algorithm, according to~\cite{qin2023deep}, the computational complexity of the actor network is $\mathcal{O}\big(\sum_{l=0}^{L_{actor}} N_{actor}^{l} N_{actor}^{l+1}\big)$, where $L_{actor}$ is the number of fully connected layers and $N_{actor}^{l}$ denotes the number of neurons in the $l$-th layer. Consequently, over $E$ episodes and $D$ decision steps, the total computational complexity of the proposed LSTM-SAC algorithm is $\mathcal{O}\big(ED\big(L_{LSTM} M_{LSTM}^{2} C_{LSTM}^{2} + \sum_{l=0}^{L_{actor}} N_{actor}^{l} N_{actor}^{l+1}\big)\big)$. This complexity scales linearly with the number of episodes $E$ and decision steps $D$, and quadratically with the network width.

\vspace{-3mm}
\section{Performance Evaluation}  \label{sec:simulation} 
\vspace{-.15mm}

In this section, we first present the simulation parameter settings and then evaluate the performance of the proposed LSTM-SAC algorithm by comparing it against three benchmark solutions under various network configurations.

\begin{figure}[!t]
\includegraphics[width=.35\textwidth]{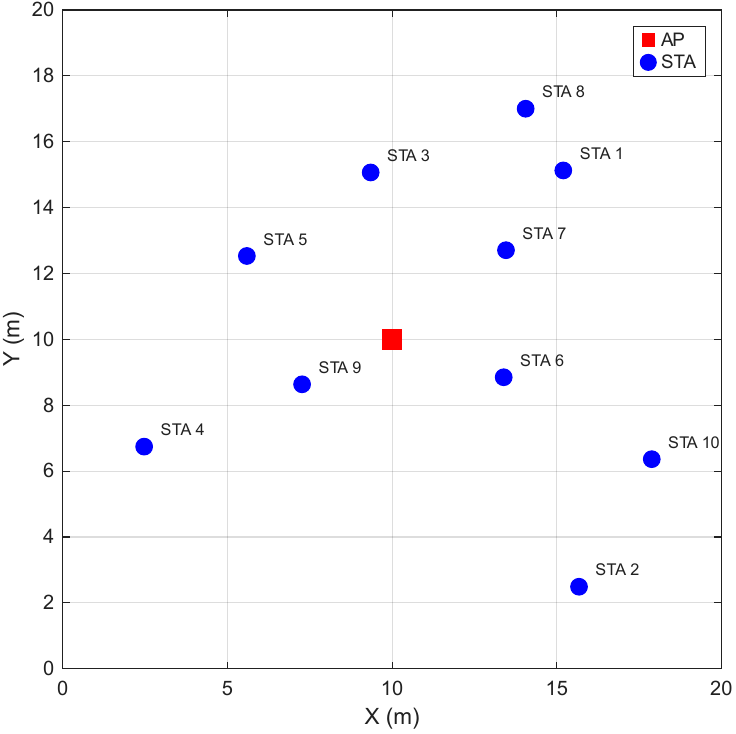}
\centering
\vspace{-1.5mm}
\caption{A Wi-Fi network consisting of one AP and 10 STAs uniformly distributed within a $20\times20$ m area.}
\label{fig:nettopology}
\end{figure}

\begin{table}[!t]
\centering
\footnotesize
\caption{MAC layer and physical layer simulation parameters ~\cite{lian2025intelligent,gao2025latency }.}
\label{tab:simulatios}
\setlength{\tabcolsep}{6pt}
\begin{tabular}{ll}
\toprule
\textbf{Parameter} & \textbf{Value} \\
\midrule
Transmission power & 20 dBm\\
Date rate in 2.4 GHz& [20, 50, 100, 150] Mbps\\
Date rate in 5 GHz & [50, 100, 200, 400] Mbps\\
Channel bandwidth & 20 MHz/40 MHz\\
Background noise & -85 dBm/-95 dBm\\
Minimum CW size ($w_{min}$) & 16\\
Maximum CW size ($w_{max}$) & 1024\\
Packet payload size & 12000 bits\\
Slot time & 9 microsecs\\
SIFS & 16 microsecs\\
AIFS & 34 microsecs\\
ACK size & 304 bits\\
Number of episodes ($E$) & 500\\
Number of decision steps ($D$) & 50\\
Reward discount factor ($\gamma$) & 0.99\\
Target entropy ($\bar{\mathcal{H}}$) & $-4N$\\
Target network update rate ($\rho$) & 0.005\\
CSMA/CA duration per decision step & 20 ms\\
\bottomrule
\end{tabular}
\end{table}

\subsection{Simulation Settings}
\vspace{-.15mm}
\subsubsection{Network Layout} We consider a $20\times20$ m square network, as shown in Fig.~\ref{fig:nettopology}, where the AP is located at the center and the STAs are uniformly distributed within the area. The AP and STAs are configured with two links operating on different frequency bands (i.e., 2.4 GHz and 5 GHz).\footnote{Although Wi-Fi 7 supports the 6 GHz band, its availability varies across regulatory regions. By contrast, the 2.4 GHz and 5 GHz bands are globally open and universally supported by Wi-Fi 7 devices. Therefore, using two links operating on 2.4 GHz and 5 GHz not only preserve the essential characteristics of MLO, but also enable us to focus on cross-layer optimization without loss of generality.} For simplicity, we assume that each band contains only one channel. The large-scale fading is modeled as $(3 \times 10^{8}/4\pi f_c\mathcal{D}(\text{STA},\text{AP}))^2$, where $f_c \in \{2.4,5\}$ GHz is the carrier frequency and $\mathcal{D}(\text{STA},\text{AP})$ denotes the Euclidean distance between a STA and the AP. The small-scale fading follows a Rayleigh distribution with scale parameter $\sqrt{1/2}$. The key simulation parameters are summarized in Table~\ref{tab:simulatios}.

\subsubsection{Algorithm Layout} We implement the LSTM-SAC algorithm using Anaconda 25.11.0 with Python 3.12.0 and PyTorch 2.9.1 on a Windows platform equipped with an Intel(R) Core(TM) i7-10700 CPU, together with an event-driven Wi-Fi simulator implemented in MATLAB R2025b. In the SAC model, the actor network consists of two fully connected (FC) hidden layers, each with 256 neurons. The critic networks are also composed of two FC hidden layers with 256 neurons each. For the LSTM model, the number of layers is set to 1, and the hidden size (i.e., the state dimension) is set to 128. The Adam optimizer is adopted, with learning rates of $10^{-3}$ and $3\times10^{-4}$ for the critic and actor networks, respectively.

\vspace{-3mm}
\subsection{Benchmark Solutions}
\vspace{-.15mm}
To demonstrate the effectiveness of the proposed LSTM-SAC algorithm, we compare its performance against three benchmark solutions:

\begin{enumerate}[leftmargin=4.5mm] 
    \item \emph{LSTM-SAC without optimizing CW sizes (LSTM-SAC w/o CW):} Following most existing Wi-Fi 7 MLO studies~\cite{alsakati2023performance,lopez2021ieee, gao2025latency, lopez2022dynamic}, LSTM-SAC w/o CW optimizes only the traffic allocation policy $\bm{\beta}(d)$ at each decision step $d$, focusing exclusively on the U-MAC layer. This baseline highlights the performance degradation that may arise from ignoring the uncertainty introduced by the CSMA/CA procedure and demonstrates the necessity of cross-layer optimization.

    \item \emph{SAC~\cite{haarnoja2018soft}:} At each decision step $d$, SAC directly takes the observation $\mathbf{o}(d)$ as input -- without using LSTM to encode historical information -- and outputs the corresponding action $\mathbf{a}(d)$. The remaining training procedure is identical to that of LSTM-SAC. This baseline is designed to illustrate the performance degradation caused by the partial observability and non-Markovian behavior inherent in Wi-Fi networks, thereby highlighting the necessity of integrating LSTM with DRL.
    \item \emph{LSTM-DDPG:} Borrowing ideas from the standard DDPG algorithm~\cite{lillicrap2015continuous}, we integrate LSTM with DDPG. DDPG extends DQN to continuous action spaces by combining deterministic policy gradients with an actor–critic architecture, enabling direct optimization of continuous actions. This baseline is designed to highlight the motivation for adopting SAC, as SAC simultaneously maximizes the expected return and the policy entropy. This entropy-regularized objective encourages persistent exploration and significantly enhances learning stability.
\end{enumerate}


\begin{figure*}[t!]
    \centering
    \begin{subfigure}[b]{.42 \textwidth}
        \includegraphics[width=\textwidth]{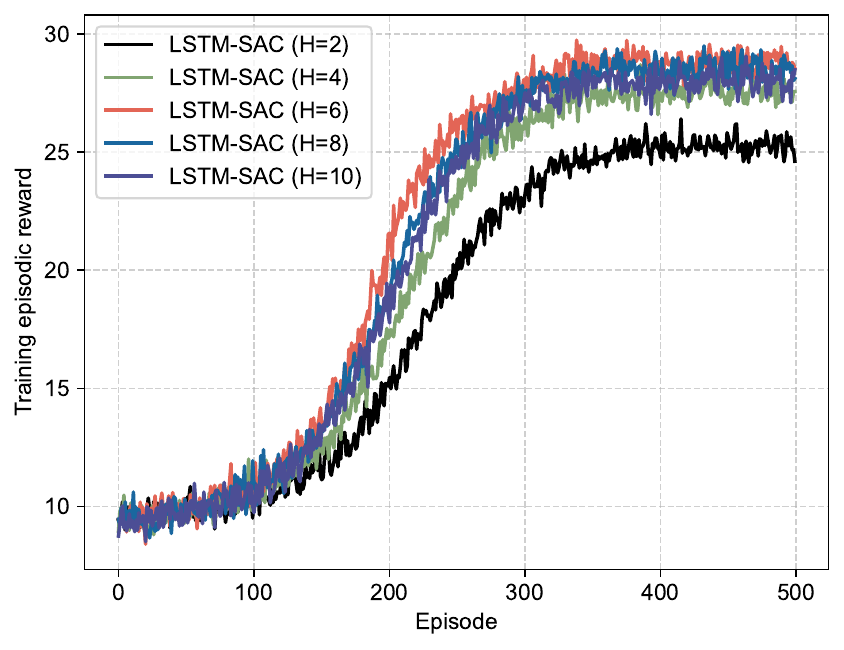}
        \vspace{-5mm}
        \caption{Impact of the number of historical observation on the reward convergence trends in LSTM-SAC (the number of STA $N=10$).}
        \label{fig:LSTMSACdifferentH}
    \end{subfigure}
    \vspace{1.5mm} 
    \begin{subfigure}[b]{.42\textwidth}
        \includegraphics[width=\textwidth]{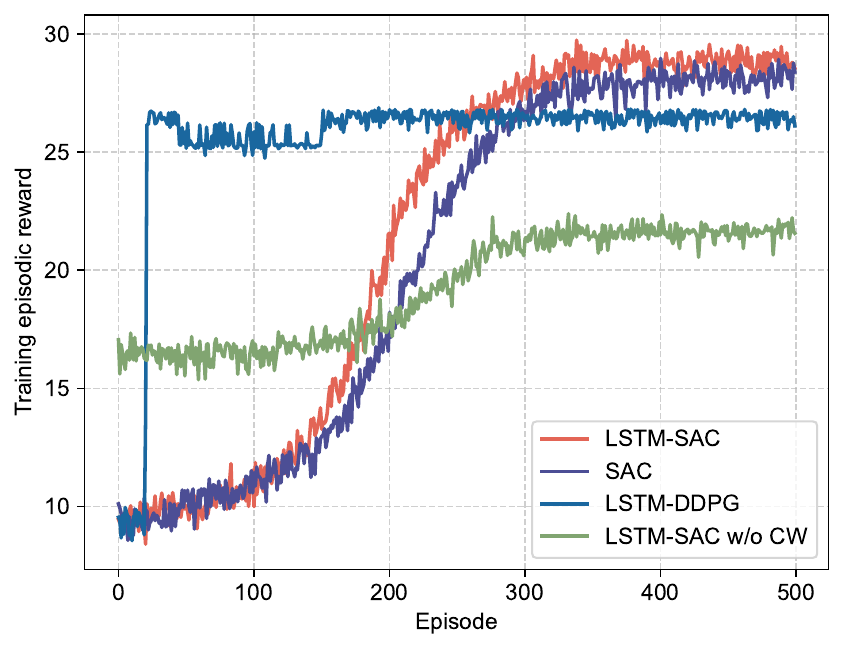}
        \vspace{-5mm}
        \caption{Comparison of reward convergence trends among different algorithms (the number of STA $N=10$).}
        \label{fig:differentALGOreward}
    \end{subfigure}
    \begin{subfigure}[b]{.42\textwidth}
        \includegraphics[width=\textwidth]{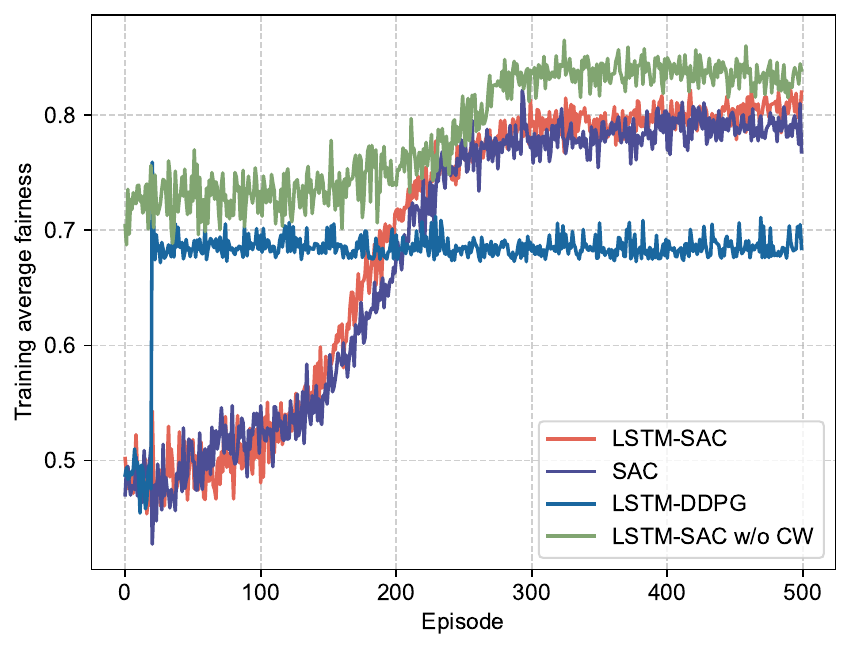}
        \vspace{-5mm}
        \caption{Comparison of fairness convergence trends among different algorithms (the number of STA $N=10$).}
        \label{fig:differentSTAnumFairness}
    \end{subfigure}
    \begin{subfigure}[b]{.43\textwidth}
        \includegraphics[width=\textwidth]{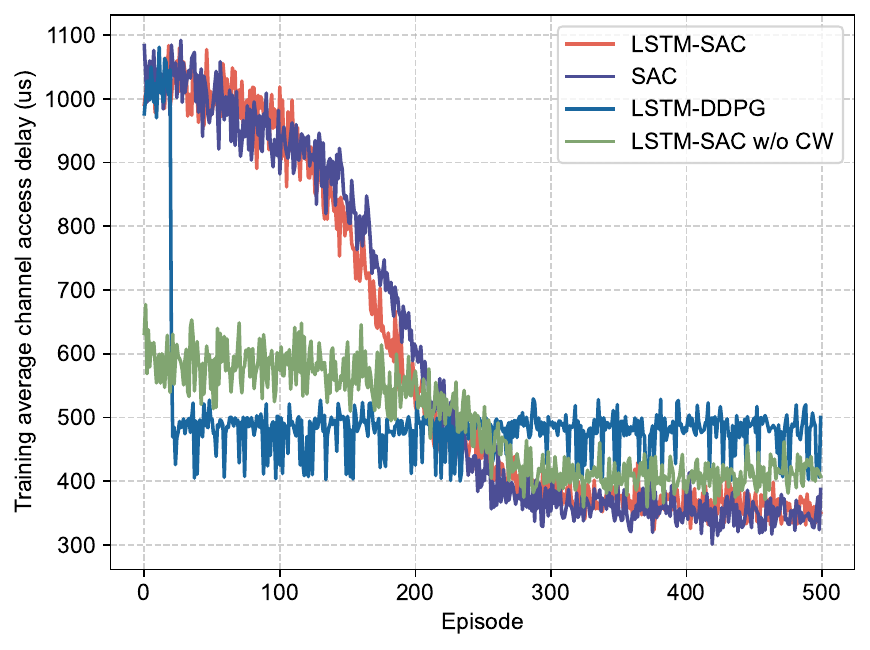}
        \vspace{-5mm}
        \caption{Comparison of channel access delay convergence trends among different algorithms (the number of STA $N=10$).}
        \label{fig:differentSTAnumAccessDelay}
    \end{subfigure}
    \vspace{-0.5mm}
    \caption{Training convergence of different algorithms under various metrics, including training reward, fairness, and channel access delay.}
    \label{fig:test}
\end{figure*}

\vspace{-3mm}
\subsection{Training Convergence Performance}
\vspace{-.15mm}

In this subsection, we analyze the convergence behavior of different algorithms during the training phase.

\subsubsection{Effect of the Numbers of Historical Observations $H$} In Fig.~\ref{fig:LSTMSACdifferentH}, we illustrate the convergence behavior of LSTM-SAC under different lengths of the historical observation window $H$. The results show that as $H$ increases from 2 to 6, the episodic reward improves, but beyond a certain point (around $H=6$), further enlarging the window leads to performance degradation. This is because a moderate history length helps the agent infer hidden system states (e.g., channel occupancy and backoff stages) and capture the non-Markovian dynamics of CSMA/CA, enabling more informed decisions. However, an overly long history introduces outdated information; due to rapidly changing traffic and contention conditions, older observations become less relevant to current decisions. Therefore, in subsequent experiments involving LSTM modules, the history window is fixed to $H=6$.\footnote{We use an LSTM rather than a Transformer because the historical observation window in our problem is short (e.g., $H=6$), for which an LSTM is sufficient to capture the relevant temporal dependencies while incurring significantly lower computational overhead. In contrast, Transformer-based architectures are typically more advantageous for modeling substantially longer sequences.}

\subsubsection{Network Throughput Training Convergence Performance} In Fig.~\ref{fig:differentALGOreward}, we illustrate the network throughput (reward) training convergence of four algorithms. The results show that LSTM-SAC consistently converges to the highest reward, followed by SAC, LSTM-DDPG, and finally LSTM-SAC w/o CW. This is because: The superiority of LSTM-SAC comes from combining LSTM-based temporal modeling with entropy-regularized SAC and cross-layer optimization. The LSTM helps infer hidden MAC-layer states under partial observability and non-Markovian CSMA/CA dynamics, while SAC ensures stable exploration. By jointly optimizing U-MAC traffic allocation and L-MAC ICW sizes, LSTM-SAC effectively balances load and controls contention, fully exploiting the throughput potential of Wi-Fi~7 MLO. SAC performs worse due to the absence of LSTM. By relying only on instantaneous observations, it fails to infer hidden state and capture non-Markovian effects caused by CSMA/CA contention and traffic reshaping. LSTM-DDPG, due to its deterministic policy gradient, lacks sufficient exploration and is therefore prone to local optima. This behavior is reflected in the reward jump, where the actor rapidly moves toward a reasonably good region once the critic provides informative gradients. LSTM-SAC w/o CW starts with a higher initial reward but converges to the lowest final throughput. Its higher initial reward stems from fixed ICW settings, whereas other methods must explore a wide range of ICW sizes and may initially select overly large ICWs that reduce early throughput. However, because LSTM-SAC w/o CW cannot adapt the ICW to changing traffic conditions, its long-term performance is limited. This observation confirms that traffic allocation alone is insufficient, as the achievable service rate is ultimately governed by CSMA/CA at the L-MAC layer, thereby motivating joint cross-layer optimization.

\subsubsection{STA Fairness Training Convergence Performance} In Fig.~\ref{fig:differentSTAnumFairness}, we illustrate the STA fairness training convergence of four algorithms using Jain’s Fairness Index $\frac{(\sum_{n=1}^N\mathcal{R}_n)^2}{N\sum_{n=1}^N(\mathcal{R}_n)^2}$, where $\mathcal{R}_n$ is the throughput of STA $n$ and values closer to 1 indicate higher fairness. The results show that LSTM-SAC, SAC, and LSTM-DDPG gradually improve fairness during training, as the agents learn more balanced traffic allocation and ICW sizes, reducing bias toward specific STAs. LSTM-SAC w/o CW achieves the highest fairness, with a slight upward trend primarily driven by the U-MAC traffic allocation. The main reason for its superior fairness is that all STAs use fixed ICWs, which impose similar channel access aggressiveness and prevent any STA from becoming overly dominant during contention. Meanwhile, allocating traffic to less congested or higher-quality links further balances the achieved throughputs, contributing to the slight upward trend. As a result, LSTM-SAC attains fairness close to LSTM-SAC w/o CW while achieving much higher throughput, showing that cross-layer optimization can coordinate traffic allocation and contention control without sacrificing fairness, thereby balancing efficiency and fairness in Wi-Fi~7 MLO networks.

\subsubsection{Channel Access Delay Training Convergence Performance} In Fig.~\ref{fig:differentSTAnumAccessDelay}, we plot the channel access delay training convergence of four algorithms. Channel access delay represents the waiting time before a successful transmission and is closely related to backoff duration and collision frequency under CSMA/CA. The results show that, for LSTM-SAC, SAC, and LSTM-DDPG, the channel access delay gradually decreases and then converges as training progresses. This is because, in the early stage of training, the agents explore a wide range of traffic allocation and ICW sizes (e.g., overly small ICWs causing collisions or overly large ICWs causing long backoff). As learning proceeds, the agents gradually discover more suitable policies that better match traffic load and contention levels, thereby reducing collisions and unnecessary waiting time. For LSTM-SAC w/o CW, the delay is also relatively stable across episodes but at a moderate level. This is because although its ICW sizes are fixed, the agent still optimizes U-MAC traffic allocation, which can slightly adjust the load distribution across links and produce mild delay variations. However, without ICW adaptation, it cannot actively regulate channel access aggressiveness under changing contention levels, so the delay reduction is limited compared to cross-layer optimization framework.

\begin{table*}[t!]
\centering
\footnotesize
\caption{Performance comparison under different numbers of STAs.}
\label{tab:overall_metrics}
\setlength{\tabcolsep}{3.5pt}
\begin{tabular}{llccccccccccc}
\toprule
\multirow{2}{*}{\textbf{Metric}} & \multirow{2}{*}{\textbf{Algorithm}} & \multicolumn{11}{c}{\textbf{Number of STAs ($N$)}} \\
\cmidrule(lr){3-13}
 &  & 5 & 6 & 7 & 8 & 9 & 10 & 11 & 12 & 13 & 14 & 15 \\
\midrule

\multirow{4}{*}{\textbf{Throughput (Gbps)}} 
& \textbf{LSTM-SAC (Ours)}        & \textbf{16.46} & \textbf{18.06} & \textbf{20.87} & \textbf{24.84} & \textbf{26.18} & \textbf{29.95} & \textbf{33.82} & \textbf{37.66} & \textbf{38.21} & \textbf{40.19} & \textbf{41.08} \\
& SAC             & 15.49 & 17.20 & 20.32 & 23.24 & 25.63 & 28.69 & 31.91 & 35.93 & 36.43 & 38.46 & 40.17 \\
& LSTM-DDPG      & 13.66 & 16.76 & 17.55 & 18.36 & 19.74 & 25.83 & 27.79 & 34.05 & 35.67 & 36.34 & 38.61 \\
& LSTM-SAC w/o CW & 11.99 & 12.59 & 15.82 & 17.97 & 18.69 & 22.19 & 25.16 & 27.78 & 28.02 & 29.44 & 31.11 \\
\midrule

\multirow{4}{*}{\textbf{Fairness}} 
& \textbf{LSTM-SAC (Ours)}        & 0.9558 & 0.8782 & 0.8564 & 0.8481 & 0.8455 & 0.8141 & 0.8531 & 0.8558 & 0.8101 & 0.8224 & 0.7914 \\
& SAC             & 0.9317 & 0.8491 & 0.8378 & 0.8004 & 0.7933 & 0.7777 & 0.8128 & 0.8348 & 0.7696 & 0.7763 & 0.7691 \\
& LSTM-DDPG       & 0.8231 & 0.7681 & 0.7632 & 0.7281 & 0.7121 & 0.6905 & 0.7335 & 0.7538 & 0.6474 & 0.6494 & 0.6257 \\
& LSTM-SAC w/o CW & \textbf{0.9652} & \textbf{0.9057} & \textbf{0.8833} & \textbf{0.8561} & \textbf{0.8533} & \textbf{0.8481} & \textbf{0.8723} & \textbf{0.8819} & \textbf{0.8131} & \textbf{0.8478} & \textbf{0.8079} \\
\midrule

\multirow{4}{*}{\textbf{Access delay (us)}} 
& \textbf{LSTM-SAC (Ours)}        & \textbf{185.69} & \textbf{275.31} & \textbf{282.44} & \textbf{299.29} & \textbf{303.76} & \textbf{319.93} & \textbf{322.59} & \textbf{352.10} & \textbf{389.79} & \textbf{390.17} & \textbf{416.54} \\
& SAC             & 216.86 & 298.16 & 305.48 & 333.35 & 334.36 & 335.50 & 339.06 & 361.31 & 403.52 & 407.39 & 423.52 \\
& LSTM-DDPG       & 286.45 & 361.48 & 368.41 & 394.91 & 403.90 & 438.72 & 442.02 & 495.88 & 506.97 & 570.17 & 696.42 \\
& LSTM-SAC w/o CW & 260.33 & 329.30 & 358.98 & 382.80 & 386.17 & 388.41 & 396.97 & 409.98 & 428.32 & 439.27 & 482.65 \\
\midrule

\multirow{4}{*}{\textbf{Running time (ms)}} 
& \textbf{LSTM-SAC (Ours)}       & 0.6651 & 0.6879 & 0.7011 & 0.7076 & 0.7129 & 0.7216 & 0.7485 & 0.7584 & 0.7755 & 0.8238 & 0.8691 \\
& SAC             & \textbf{0.2698} & \textbf{0.2838} & \textbf{0.2839} & \textbf{0.2848} & \textbf{0.2856} & \textbf{0.2905} & \textbf{0.2999} & \textbf{0.3025} & \textbf{0.3026} & \textbf{0.3034} & \textbf{0.3264} \\
& LSTM-DDPG       & 0.5787 & 0.5925 & 0.6225 & 0.6293 & 0.6307 & 0.6398 & 0.6449 & 0.6459 & 0.6606 & 0.7001 & 0.7973 \\
& LSTM-SAC w/o CW & 0.6707 & 0.6785 & 0.6904 & 0.7006 & 0.7014 & 0.7072 & 0.7861 & 0.8276 & 0.8487 & 0.8674 & 0.9044 \\
\bottomrule
\end{tabular}
\end{table*}

\vspace{-3mm}
\subsection{Verification of the Proposed LSTM-SAC Algorithm}
\vspace{-.15mm}

In this subsection, we evaluate the deployment (inference) performance of the trained models. The following results reflect the steady-state performance of the learned policies after training, where the policies are fixed and executed without exploration.

\subsubsection{Network Throughput Test Performance}
In Table~\ref{tab:overall_metrics}, we first examine the impact of the number of STAs on network throughput under different algorithms. The results show that, for all algorithms, throughput increases monotonically with the number of STAs. This is because more contending STAs inject additional traffic into the network, enabling better exploitation of multi-link transmission opportunities and improved spatial and temporal channel utilization under saturated traffic conditions. Since network throughput is directly used as the training reward, the underlying causes of the performance differences among algorithms are consistent with the reward-based analysis and are therefore not repeated here. Overall, the proposed LSTM-SAC outperforms the other algorithms, achieving performance improvements of 6.26\% over SAC, 20.49\% over LSTM-DDPG, and 37.28\% over LSTM-SAC w/o CW when the number of STAs is set to 5. Additionally, when the number of STAs is 15, LSTM-SAC outperforms SAC by 2.27\%, LSTM-DDPG by 6.39\%, and LSTM-SAC w/o CW by 32.05\%.

\subsubsection{STA Fairness Test Performance} 
We then examine the impact of the number of STAs on fairness under different algorithms. The results show that fairness varies only slightly as the number of STAs increases from 5 to 15 for all methods. This is because Jain’s Fairness Index reflects the relative throughput distribution among STAs rather than absolute throughput. Under saturated traffic and similar contention rules, increasing the number of STAs tends to scale their throughputs in a comparable manner, resulting in limited variation in fairness. Overall, LSTM-SAC w/o CW achieves the highest fairness across different STA numbers, mainly due to its fixed ICW configuration, which enforces similar channel access aggressiveness among STAs. The proposed LSTM-SAC consistently ranks second, achieving fairness close to that of LSTM-SAC w/o CW while simultaneously delivering higher throughput, thereby demonstrating a strong balance between efficiency and fairness.

\subsubsection{Channel Access Delay Test Performance} 
Also, we examine the impact of the number of STAs on channel access delay under different algorithms. The results show that, for all algorithms, channel access delay generally increases as the number of STAs grows. This is expected because more contending STAs intensify channel competition under CSMA/CA, leading to higher collision probability and longer backoff durations. As a result, packets experience longer waiting times before successful transmission. Overall, the proposed LSTM-SAC consistently achieves the lowest channel access delay across all STA numbers. This is because its cross-layer design jointly optimizes traffic allocation and ICW sizes, enabling it to balance load across links while regulating channel access aggressiveness to mitigate excessive contention. Overall, the proposed LSTM-SAC achieves the lowest channel access delay, reducing the delay by 31.17\% compared to SAC, 35.18\% compared to LSTM-DDPG, and 28.67\% compared to LSTM-SAC w/o CW when the number of STAs is set to 5. Additionally, when the number of STAs is 15, LSTM-SAC reduces the delay by 1.65\% relative to SAC, 40.18\% relative to LSTM-DDPG, and 13.69\% relative to LSTM-SAC w/o CW.


\subsubsection{Algorithm Running Time Performance} We finally examine how the number of STAs affects the algorithm running time per decision step. The results show that the running time generally increases with the number of STAs for all methods. This is because a larger STA population enlarges the state and action spaces, so the neural networks must handle higher-dimensional inputs and outputs, resulting in greater computation per forward pass. Among the compared methods, SAC achieves the lowest running time, as it relies only on feedforward networks with instantaneous observations. In contrast, other methods requires more running time due to the additional LSTM module that extracts temporal features from historical observations. The recurrent operations and extra parameters inevitably introduce computational overhead. Nevertheless, the runtime of LSTM-SAC remains at the millisecond level. Considering that LSTM-SAC achieves the highest throughput, the lowest channel access delay, and fairness comparable to LSTM-SAC w/o CW, we conclude that it delivers superior overall performance with only a modest increase in computational complexity.




\section{Conclusions and Future Works} \label{sec:conclusion}
\vspace{-.15mm}
In this paper, we proposed a novel cross-layer optimization framework for Wi-Fi 7 MLO. Specifically, we addressed the joint problem of traffic allocation at the U-MAC layer and ICW optimization at the L-MAC layer with the objective of maximizing network throughput. We first extended the classical single-link Bianchi Markov model to develop a new analytical framework for Wi-Fi 7 MLO, deriving explicit expressions that characterize throughput as functions of both the traffic allocation policy and ICW size. To effectively solve the resulting non-convex and temporally coupled optimization problem, we then introduced an LSTM-SAC algorithm that leverages LSTM networks to encode historical observations into compact latent representations, thereby mitigating the partial observability and non-Markovian behaviors inherent in Wi-Fi networks. Numerical results demonstrate that the proposed LSTM-SAC algorithm significantly outperforms existing benchmark solutions across diverse network settings.

There are several promising directions for future work. First, our analytical framework focuses on the STR mode, where links operate independently across frequency bands. Extending the cross-layer optimization to NSTR or hybrid STR/NSTR configurations would be valuable, as these modes require tight temporal coordination and introduce complex inter-link dependencies that challenge analytical modeling. Second, the current evaluation is conducted using an event-based \emph{Matlab} simulator integrated with \emph{Python}-based DRL. Real-world deployment involves additional uncertainties such as imperfect carrier sensing and device heterogeneity. Therefore, validating the proposed framework on real Wi-Fi 7 testbeds or FPGA-based prototypes~\cite{ding2019gplm} is an important future step. Such hardware-in-the-loop experiments would help close the gap between simulation and practice and offer deeper insights into the feasibility, scalability, and robustness of intelligent cross-layer optimization for Wi-Fi 7 MLO networks.
\vspace{-3mm}


\bibliographystyle{IEEEtran}

\vspace{-14mm}

\end{document}